\documentclass[a4paper,12pt]{article}

\usepackage{psfig}

\newcommand{\resetequ}{\setcounter{equation}{0}}

\usepackage{amssymb}

\def\bone{{\bf 1}}

\def\bbbone{{\mathchoice {\rm 1\mskip-4mu l} {\rm 1\mskip-4mu l}
{\rm 1\mskip-4.5mu l} {\rm 1\mskip-5mu l}}}
\def\NN{{\mathchoice {\mathrm{I \hspace{-0.2em} N}}
    {\mathrm{I \hspace{-0.2em} N}} {\mathrm{I \hspace{-0.14em} N}}
    {\mathrm{I \hspace{-0.14em} N}}}}
\def\ZZ{{\mathchoice {\mathsf{Z \hspace{-0.45em} Z}} {\mathsf{Z
        \hspace{-0.45em} Z}} {\mathsf{Z \hspace{-0.32em} Z}}
    {\mathsf{Z \hspace{-0.23em} Z}}}}

\def\RR{{\mathchoice {\mathrm{I \hspace{-0.2em} R}} {\mathrm{I
        \hspace{-0.2em} R}} {\mathrm{I \hspace{-0.14em} R}}
    {\mathrm{I \hspace{-0.14em} R}}}}

\newfam\gotfam
\font\twlgot=eufm10 at 12pt \font\tengot=eufm10
 \font\sevengot=eufm7
\textfont\gotfam=\twlgot \scriptfont\gotfam=\tengot
\scriptscriptfont\gotfam=\sevengot
\def\got{\fam\gotfam\twlgot}

\newtheorem{prop}  {Proposition}
\newtheorem{lemma}  {Lemma}
\newtheorem{theor}   {Theorem}

\newcommand{\be}  {\begin{equation}}
\newcommand{\ee}  {\end{equation}}
\newcommand{\bea} {\begin{eqnarray}}
\newcommand{\eea} {\end{eqnarray}}
\newcommand{\lp}  {\left(}
\newcommand{\rp}  {\right)}

\newcommand{\Br}  {\overline}

\newcommand{\cB}  {{\cal B}}
\newcommand{\cL}  {{\cal L}}

\newcommand{\cC}  {{\cal C}}
\newcommand{\cD}  {{\cal D}}

\newcommand{\cM}  {{\cal M}}
\newcommand{\cS}  {{\cal S}}

\newcommand{\cI}  {{\cal I}}

\newcommand{\cR}  {{\cal R}}
\newcommand{\cA}  {{\cal A}}

\newcommand{\om}  {\omega}

\newcommand{\si}  {\sigma}
\newcommand{\ga}  {\gamma}
\newcommand{\Ga}  {\Gamma}

\newcommand{\al}  {\alpha}
\newcommand{\la}  {\lambda}
\newcommand{\rh}  {\rho}

\newcommand{\La}  {\Lambda}

\newcommand{\de}  {\delta}
\newcommand{\De}  {\Delta}

\newcommand{\Ph}  {\Phi}
\newcommand{\ph}  {\phi}

\newcommand{\Gg}  {{\got g}}

\newcommand{\und}  {\underline}
\newcommand{\til}  {\tilde}

\def\Br{\overline}

\newcommand{\eqdef} {\stackrel{\rm def}{=}}

\def\endproof{\hfill\vrule height .6em width .6em depth
  0pt\goodbreak\vskip.25in}

\begin{document}

\title{Bosonic Monocluster Expansion}

\author{A. Abdesselam\\ {\small D{\'e}partement de Math{\'e}matiques}\\
{\small  Universit{\'e} Paris XIII, Paris-Nord, Villetaneuse}\\
{\small Avenue J.B. Cl{\'e}ment, F93430 Villetanneuse, France}\\
\\
J. Magnen and V. Rivasseau\\
{\small 
Centre de physique th{\'e}orique, CNRS UMR 7644}\\
{\small Ecole Polytechnique, F91128 Palaiseau Cedex, France}
}

\maketitle

\hfill\eject

{\abstract{We compute connected Green's
functions of a Bosonic field theory
with cutoffs by means of a ``minimal'' expansion which 
in a single move, interpolating a generalized propagator,
performs the usual tasks of the cluster
and Mayer expansion. In this way it allows a direct construction 
of the infinite volume or thermodynamic limit and it brings constructive
Bosonic expansions closer to constructive Fermionic 
expansions and to perturbation theory.} }

 
\hfill\eject

\section{Introduction}

A key problem in physics is to construct the thermodynamic limit of
large systems. Only intensive or normalized quantities have a well defined
limit. For a Bosonic field theory the standard way to construct this 
limit is to introduce first a finite volume cutoff, then to perform 
a cluster expansion, which writes the theory as a polymer gas but with hardcore
constraints, then to perform a Mayer expansion which removes these constraints
by comparing this gas to a perfect gas [9]. 
It is still
slightly frustrating for two reasons.

Firstly for Fermionic theories there is no need of such 
a sequence of two expansions on top of each other: 
a single tree formula expresses directly the
infinite volume limit of normalized functions as a convergent series [3].
It is therefore desirable to have such a single formula computing
directly the infinite volume limit of connected Green's functions in the
Bosonic case too. 

Secondly mathematically both the cluster and the Mayer expansions
can be written elegantly using forest formulas [1]; they have therefore
some common nature, which led us to suspect for quite a while that
there should exist a single expansion performing both tasks
at the same time. In fact the first example of such
a formula was given in [1], but it is still really a somewhat
artificial mixing of the two expansions (using a two
stages formula technically called a "jungle" formula), and
it is not obtained by interpolating propagators only. 

In this paper we propose a much more natural solution to this problem, which
writes directly the infinite volume limit of normalized functions as a 
convergent series. The Mayer expansion can be understood as taking
place in some extended space of copies. Therefore
we propose, for any space $\RR^{d}$, to define the Mayer space as 
$\RR^{d} \times \NN$. In this extended space we introduce 
expansions steps which interpolate solely the (generalized) propagator
of the extended theory.
The outcome of our expansion is not exactly but almost a tree formula
in this extended Mayer space-time. It generates a single cluster
(hence we name our expansion a ``monocluster'' expansion), 
and the profile of this
cluster in the Mayer space is a solid-on-solid profile, with no
overhangs. This means that our expansion 
makes truly a minimal use of the Mayer copies. 

We hope to extend this analysis in the future to multiscale
expansions such as the one of [2], written for
the infrared $\phi^{4}_{4}$ model. This would suppress the 
need for iteration of Mayer expansions to perform renormalization (probably
the most cumbersome aspect of explicit multiscale expansions). In this
way we hope to obtain a completely
explicit non-perturbative solution of the renormalization
group induction for Bosonic theories (apart from the inductive computation
of the effective constants). It would bring these Bosonic 
theories to the same level of understanding than Fermionic 
theories, for which such explicit solutions are known [5].
For a review of rigorous renormalization group methods for bosonic field
theory models we refer the reader to [4, 6, 7, 10].

\section{The Model}
\resetequ

Let $C(x,y)$ be the smooth translation-invariant kernel of a
covariance operator on $\RR^d$, i.e. such that $(f,g)\mapsto
<f,Cg>_{L^2(\RR^d)}$ is a positive continuous bilinear form on the
Schwartz space $\cS(\RR^d)$. By the Bochner-Minlos theorem
(see [8]), there is an associated Gaussian measure $d\mu_C$ on
$\cS'(\RR^d)$ with covariance $C$. The smoothness of $C$ insures
that $d\mu_C$ is supported on smooth functions.

We assume that $C$ satisfies a condition of rapid decay:
\be
\forall r\ge 1,
\exists K_1(r)>0,
\forall x,y\in\RR^d,
|C(x,y)|\le
K_1(r) (1+|x-y|)^{-r}
\label{rapdec}
\ee
Let $P(x)$ be a real polynomial with even degree $2m$ and positive
leading coefficient.
There is then a constant $K_2>0$ such that, for all $x\in\RR$,
$|P(x)|\le K_2(1+x^{2m})$.
We introduce a discretization
\be
\cD\eqdef\left\{\prod_{i=1}^d [k_i,k_i+1[\ |
\ (k_1,\ldots,k_d)\in\ZZ^d \right\}
\ee
of $\RR^d$ with boxes $\De$ of unit
size. If $x\in\RR^d$, we denote by $\De(x)$ the unique $\De\in\cD$
containing $x$. We denote by $\La$ a hypercube of $\RR^d$ that is
a union of boxes in $\cD$, and by $|\La|$ the number of these
boxes, which also happens to be equal to $vol(\La)$.

For any $\la\ge 0$, we introduce a partition function with free
boundary conditions:
\be
Z(\La)=
\int d\mu_C(\ph)
\exp\lp
-\la\int_\La P(\ph(x))dx
\rp
\label{zlambda}
\ee
as well as unnormalized Schwinger functions, for $x_1,\ldots,x_n$
in $\RR^d$:
\be
S_{\La,u}(x_1,\ldots,x_n)\eqdef
\int d\mu_C(\ph)
\ph(x_1)\cdots\ph(x_n)
\exp\lp
-\la\int_\La P(\ph(x))dx
\rp
\ee
These are well defined quantities, besides $Z(\La)>0$.
Indeed, by Jensen's inequality and Wick's theorem (see [8]),
\bea
Z(\La) & \ge & \exp\lp\int d\mu_C(\ph) (-\la)\int_\La P(\ph(x))dx
\rp \\
 & \ge & \exp\lp -K_2\la\int_{\La}dx\int d\mu_C(\ph)
 (1+\ph(x)^{2m})\rp \\
 & \ge & \exp\lp -K_2\la|\La|\lp 1+\frac{(2m)!}{2^m m!} C(0,0)\rp\rp > 0
\ \ .
\eea
One can thus consider the finite-volume {\it normalized} Schwinger
functions, or correlation functions,
\be
S_{\La}(x_1,\ldots,x_n)\eqdef
\frac{S_{\La,u}(x_1,\ldots,x_n)}{Z(\La)}
\ee
and study their thermodynamic limit when $\La\nearrow\RR^d$.

The typical example we have in mind is the $\ph^4$ theory in
a single slice of momenta, that is with both ultraviolet and
infrared cut-offs as defined e.g. by the choice:
\be
C(x,y)\eqdef \int\frac{d^d p}{(2\pi)^d} e^{ip(x-y)}\frac{e^{-p^2}}{p^2+1}
\ee
and $P(x)=x^4$.

One of the classical results we rederive using our new expansion
scheme is
\begin{theor}
There exists $\la_0>0$, such that, for any $\la\in [0,\la_0]$,
any $n\ge 1$, and $x_1, \ldots,x_n\in\RR^d$,
$S(x_1,\ldots,x_n)=\lim_{\La\nearrow\RR^d}S_{\La}(x_1,\ldots,x_n)$
exists.
\end{theor}

Of course, more results can be obtained with our method, like
Borel summability of perturbation theory, or complete asymptotic
expansion of the decay rate of $S(x_1,x_2)$ etc\ldots
But as explained in the introduction,
our purpose here is rather to present, at work, a new
expansion scheme in the cluster expansion business that produces a
sum over a single polymer (i.e. set of cubes), and therefore
completely avoids the so-called Mayer expansion.

\section{The expansion}
\resetequ

We first introduce a denumerable set of copies of the field $\ph$.
We let $\cL\eqdef \cD\times\NN$ which we identify with a
discretization of the ``Mayer space'' 
$\RR^d\times\NN$. For $\cM$ a positive matrix
with entries indexed by elements $b$ of $\cL$, we define the
covariance operator on $\RR^d\times\NN$:
\be
\cC[\cM](x,k;x'k')=
C(x,x')\cM(b(x,k),b(x',k'))
\ee
where $b(x,k)=(\De(x),k)$
denotes, with a slight abuse of terminology, the box of $\cL$
containing the pair $(x,k)$.
In particular we consider $\cM_\emptyset$ defined by
\be
\cM_\emptyset((\De,k);(\De',k'))=
\left\{
\begin{array}{ll}
1 & {\rm if}\ k=k'=0\\
\de_{\De,\De'} & {\rm if}\ k=k'\ge 1\\
0 & {\rm otherwise}
\end{array}
\right.
\ee
i.e. in block form
\be
\cM_\emptyset=
\begin{array}{c}
\begin{array}{rc}
\cL_0 & \cL_{\ge 1}
\end{array}\\
\lp
\begin{array}{cc}
\bone & 0\\
0 & {\rm Id}
\end{array}
\rp\\
\;
\end{array}
\ee
where $\cL_0\eqdef\cD\times\{0\}$, $\cL_{\ge1}\eqdef
\cD\times\NN^\ast$, $\bone$ is the matrix with entries 1
everywhere and ${\rm Id}$ is the identity matrix.
Clearly, $\cC_\emptyset=\cC[\cM_\emptyset]$ is a positive
covariance operator; and we can define $d\mu_{\cC_\emptyset}(\Ph)$
the measure of a Gaussian random field $\Ph(x,k)$ on
$\RR^d\times\NN$, with covariance $\cC_\emptyset$.
We introduce also the notations $\cD_\La\eqdef\{\De\in\cD|
\De\subset\La\}$, and for any integer $N\ge 0$,
$\cL_{\La,N}\eqdef\cD\times\{0,1,\ldots,N\}\subset\cL$.

Now consider
\bea
\lefteqn{
H_{\La,N}(x_1,\ldots,x_n)\eqdef} & & \nonumber\\
 & & \int d\mu_{\cC_\emptyset}(\Ph)
\prod_{i=1}^{n}
\Ph(x_i,0)
\exp\lp -\la\sum_{(\De,k)\in\cL_{\La,N}}\int_\De P(\Ph(x,k))dx
\rp \ \ .
\eea
We obviously have, due to the definition of $\cC_\emptyset$,
the factorization
\be
H_{\La,N}(x_1,\ldots,x_n)=S_{\La,N}(x_1,\ldots,x_n)
\cdot
Z_0^{N|\La|}
\ee
where
\be
Z_0\eqdef
\int d\mu_{\bbbone_{\De}C\bbbone_{\De}}
\exp\lp -\la\int_\De P(\Ph(x,k))dx
\rp
\ee
the normalization of an isolated cube, does not depend on $\De$,
since the kernel $C$ is translation-invariant. Here, $\bbbone_\De$
denotes the sharp characteristic function of $\De$. Note that
$Z_0$ differs from $Z(\De)$ by a choice of boundary condition.
We now proceed to write an expansion for
$H_{\La,N}(x_1,\ldots,x_n)$, after introducing some combinatorial
definitions.

First we define the notion of a \und{polymer}. We let $\Ga_0\eqdef
\{\De\in\cD|\exists i, x_i\in\De\}\times\{0\}\subset\cL_0$.
We also define $\Ga_{-1}\eqdef\emptyset$. We then say that a
finite set $\Ga\subset\cL$ is polymer if, whenever $(\De,k)\in\Ga$,
we also have $(\De,k')\in\Ga$ for any $k'$, $0\le k'\le k$. We
also introduce the \und{altitude} function $h_\Ga$ of a polymer,
on $\cD$ as:
\be
h_\Ga(\De)\eqdef\left\{
\begin{array}{ll}
-1 & {\rm if}\ \{k|(\De,k)\in\Ga\} = \emptyset\\
\max \{k|(\De,k)\in\Ga\} & {\rm otherwise.}
\end{array}
\right.
\ee
A polymer $\Ga$ is uniquely determined by its altitude function
$h_\Ga$. We also introduce the \und{roof} $W(\Ga)\subset\cL$ of a
polymer $\Ga$ as:
\be
W(\Ga)\eqdef\{(\De,h_\Ga(\De)+1)|\De\in\cD\}
\ee
and its \und{sky} $S(\Ga)\eqdef\cL\backslash(\Ga\cup W(\Ga))$.
The sets $\Ga$, $W(\Ga)$ and $S(\Ga)$ then form a partition of
$\cL$.

Let $\Gg=(l_1,\ldots,l_p)$ be an ordered sequence of unordered
pairs of the form $l=\{b,b'\}$ with $b$, $b'$ distinct elements of
$\cL$. $p=0$ corresponding to $\Gg=\emptyset$ is allowed too. We
define, for $1\le i\le p$, $\Ga_{i,\Gg}\eqdef\Ga_0\cup l_1\cup
\cdots\cup l_i$. We also set, by convention, $\Ga_{0,\Gg}\eqdef
\Ga_0$ and $\Ga_{-1,\Gg}\eqdef\Ga_{-1}=\emptyset$.
We say that $\Gg$ is a \und{cluster-graph} if, for any $i$,
$1\le i \le p$, the unordered pair $l_i$ is of the form $\{b,b'\}$
for some $b$ and $b'$ that satisfy one of the following two
conditions:

(i) $b\in\Ga_{i-1,\Gg}$ and $b'\in W(\Ga_{i-1,\Gg})$

(ii) $b$, $b'\in W(\Ga_{i-1,\Gg})$ and $b\notin\cL_0$.

It is easy to check that $\Ga_{i,\Gg}$ defined previously is
indeed a polymer, for any $i$, $1\le i \le p$.
A pair $l_i$, which is called a \und{link} of the graph $\Gg$, is
said of type {\em cluster-roof} or $\Ga W$ if (i) occurs, and of
type {\em roof-roof} or $WW$ if (ii) occurs (see Fig.1).

\begin{figure}
\psfig{figure=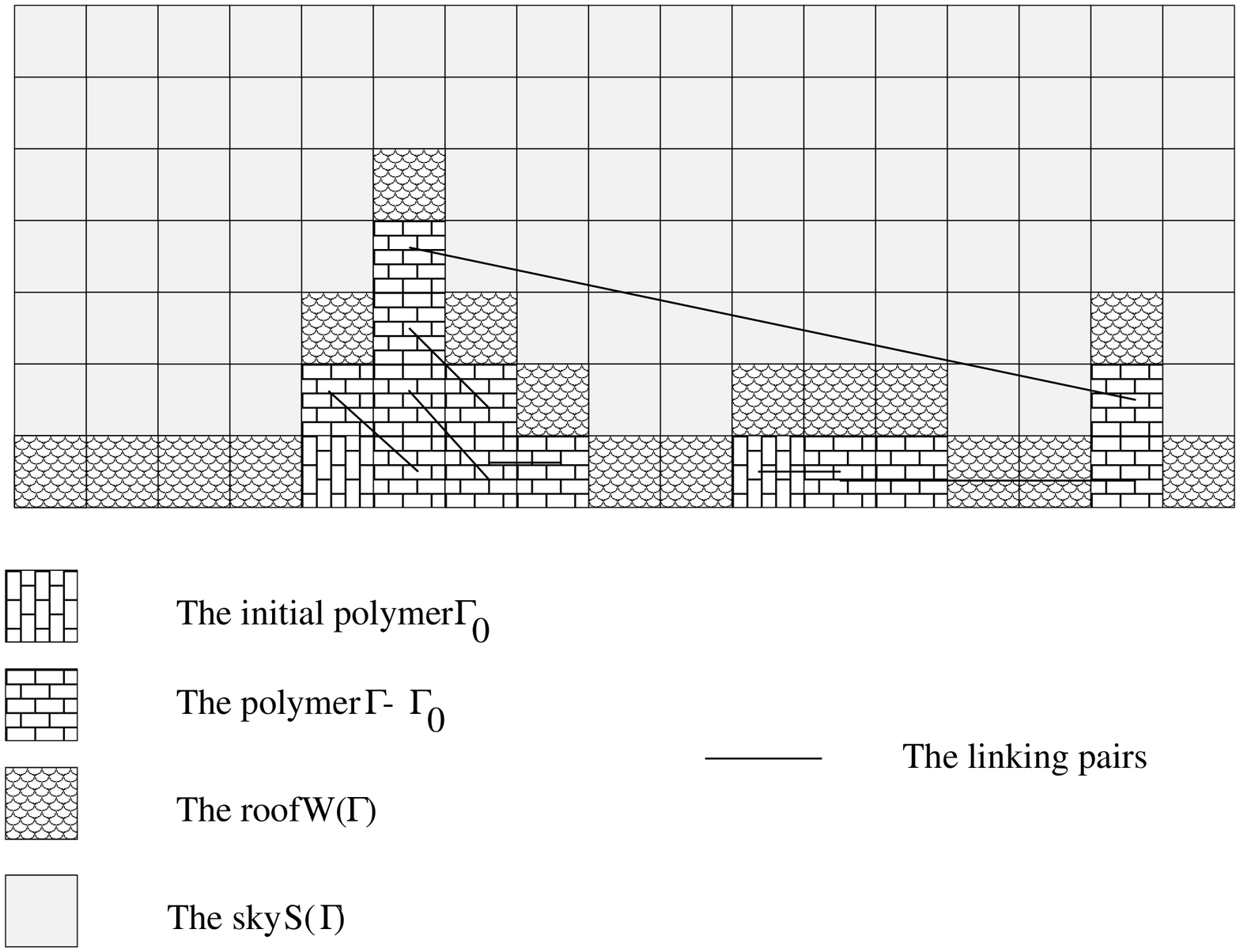,width=13cm}
\caption{A cluster graph}
\end{figure}

If $b\in\cL$, we define the \und{conception index} of $b$ with
respect to $\Gg$:
\be
\mu_\Gg(b)\eqdef
\inf\lp
\{ i| -1\le i\le p, b\in W(\Ga_{i,\Gg})\}\cup\{p+1\}
\rp
\label{mu}
\ee
and the \und{creation index} of $b$:
\be
\nu_\Gg(b)\eqdef
\inf\lp
\{ i| -1\le i\le p, b\in \Ga_{i,\Gg}\}\cup\{p+1\}
\rp
\label{nu}
\ee
Note that we always have $\mu_\Gg(b)<\nu_\Gg(b)$ if $b\in
(\Ga_{p,\Gg}\cup W(\Ga_{p,\Gg}))$. Indeed, by definition of a
cluster-graph $\Ga_{i,\Gg}\backslash\Ga_{i-1,\Gg}=l_i\backslash
\Ga_{i-1,\Gg}\subset W(\Ga_{i-1,\Gg})$.
In fact, $W(\Ga_{i})$ can be viewed as a solid-on-solid interface
that elevates in $\cL$ as the cluster $\Ga_{i,\Gg}$ grows with
increasing $i$. A cube $b$ has to belong to a $W(\Ga_{i,\Gg})$
before it belongs to a $\Ga_{i,\Gg}$.
If $b$, $b'$ are two elements of $\cL$ we let:
\be
s\mu_\Gg(b,b')\eqdef
\max(\mu_\Gg(b),\mu_\Gg(b'))
\ee
\be
s\nu_\Gg(b,b')\eqdef
\max(\nu_\Gg(b),\nu_\Gg(b'))
\ee
and
\be
i\nu_\Gg(b,b')\eqdef
\min(\nu_\Gg(b),\nu_\Gg(b'))\ \ .
\ee

Now given a decreasing vector ${\bf h}$ of $p+1$ parameters
$1>h_1>\cdots>h_p>h_{p+1}>0$ with the additional convention
$h_0\eqdef 1$ and $h_{-1}\eqdef +\infty$ so that
$\frac{1}{h_{-1}}=0$, we define the following matrix
$\cM_{\Gg,{\bf h}}$ on $\cL$.
For $b$, $b'$ in $\cL$ we let
\be
\cM_{\Gg,{\bf h}}\eqdef
\left\{
\begin{array}{ll}
1 & {\rm if}\ b=b'\\
0 & {\rm if}\ b\neq b'\ {\rm and}\ s\mu_\Gg(b,b')\ge
i\nu_\Gg(b,b')\\
h_{s\nu_\Gg(b,b')}\lp\frac{1}{h_{i\nu_\Gg(b,b')}}
-\frac{1}{h_{s\mu_\Gg(b,b')}}\rp &
{\rm if}\ b\neq b'\ {\rm and}\ s\mu_\Gg(b,b')<
i\nu_\Gg(b,b').
\end{array}
\right.
\label{covint}
\ee
We will later prove that $\cM_{\Gg,{\bf h}}$ is a positive matrix.
Before that, we introduce the following operation on covariance
matrices on $\cL$.
If $\Ga$ is a polymer, and $\cM$ is a matrix on $\cL$, we define
the new matrix $T_\Ga[\cM]$ by
\be
T_\Ga[\cM](b,b')\eqdef
\left\{
\begin{array}{ll}
\cM(b,b') &  {\rm if}\ b,b'\in\Ga\\
1 &  {\rm if}\ b,b'\in W(\Ga)\\
\de_{b,b'} & {\rm if}\ b,b'\in S(\Ga)\\
0 & {\rm otherwise}
\end{array}
\right.
\label{tronc}
\ee
or in block form
\be
T_\Ga(\cM)=
\begin{array}{c}
\begin{array}{rcl}
\ \ \ \ \Ga & W(\Ga) & S(\Ga)
\end{array}\\
\lp
\begin{array}{ccc}
\cM {|}_{\Ga} & 0 & 0\\
0 & \bone & 0\\
0 & 0 & {\rm Id}
\end{array}
\rp\\
\;
\end{array}\ \ \ .
\ee
Obviously $T_\Ga[\cM]$ is positive if $\cM$ is.
\begin{lemma}
If $\Gg=(\Gg',l_p)$ is a cluster-graph of length $p\ge 1$, and
${\bf h}=({\bf h}',h_{p+1})$ is a decreasing vector of parameters,
we have
\be
\cM_{\Gg,{\bf h}}=
\frac{h_{p+1}}{h_p}\cM_{\Gg',{\bf h}'}+
\lp 1-\frac{h_{p+1}}{h_p}\rp
T_{\Ga_{p,\Gg}}[\cM_{\Gg',{\bf h}'}]
\label{deccov}
\ee
\end{lemma}
\noindent{\bf Proof :}
We check the equality for every pair of boxes $b$, $b'$ in $\cL$.
The case $b=b'$ holds trivially.

$\bullet$ If $b\neq b'$ are both in $\Ga_{p,\Gg}$, then the choice of upper
cut-off on the infimum in (\ref{mu}) and (\ref{nu}) readily implies
that $\mu_\Gg(b)=\mu_{\Gg'}(b)\le p$ and $\nu_\Gg(b)=
\nu_{\Gg'}(b)\le p$. Therefore, 
\be
\cM_{\Gg,{\bf h}}(b,b')=
\cM_{\Gg',{\bf h}'}(b,b')=
T_{\Ga_{p,\Gg}}[\cM_{\Gg',{\bf h}'}](b,b')
\ee
so that (\ref{deccov}) holds.

$\bullet$ If $b\neq b'$ are both in $W(\Ga_{p,\Gg})$, then
$\mu_\Gg(b)=\mu_{\Gg'}(b)\le p$ whereas $\nu_\Gg(b)=p+1$,
$\nu_{\Gg'}(b)=p$ and likewise for $b'$.
Therefore
\be
\cM_{\Gg,{\bf h}}(b,b')=
h_{p+1}\lp\frac{1}{h_{p+1}}-\frac{1}{h_{s\mu_{\Gg'}(b,b')}}\rp
\ee
\be
\cM_{\Gg',{\bf h}'}(b,b')=
h_{p}\lp\frac{1}{h_{p}}-\frac{1}{h_{s\mu_{\Gg'}(b,b')}}\rp
\ee
whereas
$T_{\Ga_{p,\Gg}}[\cM_{\Gg',{\bf h}'}](b,b')=1$,
and thus
\bea
\lefteqn{
\frac{h_{p+1}}{h_p}\cM_{\Gg',{\bf h}'}(b,b')+
\lp 1-\frac{h_{p+1}}{h_p}\rp
T_{\Ga_{p,\Gg}}[\cM_{\Gg',{\bf h}'}](b,b')} &  & \nonumber \\
 & & =\frac{h_{p+1}}{h_p}
h_p\lp\frac{1}{h_{p}}-\frac{1}{h_{s\mu_{\Gg'}(b,b')}}\rp
+ \lp 1-\frac{h_{p+1}}{h_p}\rp \\
 &  & = h_{p+1}\lp\frac{1}{h_{p+1}}-
\frac{1}{h_{s\mu_{\Gg'}(b,b')}}\rp
\eea
so that (\ref{deccov}) holds.

$\bullet$ If $b\in\Ga_{p,\Gg}$ and $b'\in W(\Ga_{p,\Gg})$,
then $\mu_\Gg(b)=\mu_{\Gg'}(b)\le p$,
$\nu_\Gg(b)=\nu_{\Gg'}(b)\le p$,
$\mu_\Gg(b')=\mu_{\Gg'}(b')\le p$, $\nu_\Gg(b')=p+1$ and
$\nu_{\Gg'}(b')=p$.
Therefore $s\mu_\Gg(b,b')=s\mu_{\Gg'}(b,b')$ and
$i\nu_\Gg(b,b')=i\nu_{\Gg'}(b,b')$.
Besides $T_{\Ga_{p,\Gg}}[\cM_{\Gg',{\bf h}'}](b,b')=0$.
So if $s\mu_\Gg(b,b')\ge i\nu_{\Gg}(b,b')$
both sides of (\ref{deccov}) vanish; else we have
\be
\cM_{\Gg,{\bf h}}(b,b')=
h_{p+1}\lp\frac{1}{h_{i\nu_{\Gg'}(b,b')}}
-\frac{1}{h_{s\mu_{\Gg'}(b,b')}}\rp
\ee
and
\be
\cM_{\Gg',{\bf h}'}(b,b')=
h_p \lp\frac{1}{h_{i\nu_{\Gg'}(b,b')}}
-\frac{1}{h_{s\mu_{\Gg'}(b,b')}}\rp
\ee
which implies (\ref{deccov}).

$\bullet$ Finally if $b\in S(\Ga_{p,\Gg})\subset S(\Ga_{p-1,\Gg'})$
and $b'\neq b$ is anywhere in $\cL$, we have
$T_{\Ga_{p,\Gg}}[\cM_{\Gg',{\bf h}'}](b,b')=0$,
$\mu_\Gg(b)=\nu_\Gg(b)=p+1$ and
$\mu_{\Gg'}(b)=\nu_{\Gg'}(b)=p$.
Thus $s\mu_\Gg(b,b')\ge i\nu_{\Gg}(b,b')$ and
$s\mu_{\Gg'}(b,b')\ge i\nu_{\Gg'}(b,b')$ so that both sides of
(\ref{deccov}) vanish again.

This completes the check in every case. \endproof

\begin{lemma}
For any cluster-graph $\Gg$ of length $p\ge 0$ and associated
decreasing parameter vector ${\bf h}$ of length $p+1$, the matrix
$\cM_{\Gg,{\bf h}}$ is positive.
\end{lemma}

\noindent{\bf Proof :}
Convex combinations and the operation $\cM\mapsto T_\Ga[\cM]$
preserve positivity; so, by induction thanks to the previous
lemma, we only need to check the $p=0$ situation.
But then $\Gg=\emptyset$, ${\bf h}=(h_1)$, and for $b\in\cL$ we
have
\be
\mu_\emptyset(b)=
\left\{
\begin{array}{ll}
-1 & {\rm if}\ b\in\cL_0\\
0  & {\rm if}\ b\in(W(\Ga_0)\backslash\cL_0)
\subset\cD\times\{1\}\\
1 & {\rm if}\ b\in S(\Ga_0)
\end{array}
\right.
\ee
and
\be
\nu_\emptyset(b)=
\left\{
\begin{array}{ll}
0 & {\rm if}\ b\in\Ga_0\\
1 & {\rm if}\ b\in W(\Ga_0)\cup S(\Ga_0)\ .
\end{array}
\right.
\ee
Now a straight-forward calculation using (\ref{covint}) show that,
in block form, we have
\be
\cM_{\emptyset, (h_1)}=
\begin{array}{c}
\begin{array}{cccr}
\ \ \Ga_0 & \ W(\Ga_0)\cap\cL_0 & W(\Ga_0)
\backslash\cL_0 & \ \ \ \ \ S(\Ga_0)
\end{array}\\
\lp
\begin{array}{cccc}
\bone & h_1\bone & 0 & 0\\
h_1\bone & \bone & (1-h_1)\bone & 0\\
0 & (1-h_1)\bone & (1-h_1)\bone+h_1 {\rm Id} & 0\\
0 & 0 & 0 & {\rm Id}
\end{array}
\rp\\
\;
\end{array}
\ee
i.e.
\be
\cM_{\emptyset, (h_1)}=
h_1
\lp
\begin{array}{cccc}
\bone & \bone & 0 & 0\\
\bone & \bone & 0 & 0\\
0 & 0 & {\rm Id} & 0\\
0 & 0 & 0 & {\rm Id}
\end{array}
\rp
+(1-h_1)
\lp
\begin{array}{cccc}
\bone & 0 & 0 & 0\\
0 & \bone & \bone & 0\\
0 & \bone & \bone & 0\\
0 & 0 & 0 & {\rm Id} 
\end{array}
\rp
\ee
or
\be
\cM_{\emptyset, (h_1)}=h_1\cM_\emptyset+(1-h_1)
T_{\Ga_0}[\cM_\emptyset]
\label{inter1}
\ee
which is clearly positive.
\endproof

Remark that we have showed, {\em en passant}, that (\ref{deccov})
really starts at $p=0$, $\cM_\emptyset$ being the matrix
corresponding to a cluster-graph of ``length -1''.
 We need some more notation to proceed. Here
$\Gg=(l_1,\ldots,l_p)$, $p\ge 0$, is a cluster-graph, ${\bf h}=
(h_1,\ldots,h_{p+1})$ is a decreasing vector of parameters.
For any $b\in\cL$, and any $\al$, $0\le \al\le p+1$, we let
\be
\mu_{\Gg,\al}(b)\eqdef
\inf\lp
\{ i| -1\le i\le \al-1, b\in W(\Ga_{i,\Gg})\}\cup\{\al\}
\rp
\ee
and
\be
\nu_{\Gg,\al}(b)\eqdef
\inf\lp
\{ i| -1\le i\le \al-1, b\in \Ga_{i,\Gg}\}\cup\{\al\}
\rp\ .
\ee
This is the same as the previously defined $\mu_\Gg(b)$ and
$\nu_\Gg(b)$, using the truncation $(l_1,\ldots,l_{\al-1})$
of $\Gg$ instead of the full graph $\Gg$.
We also denote for $b$, $b'$ in $\cL$,
\be
s\mu_{\Gg,\al}(b,b')\eqdef\max(\mu_{\Gg,\al}(b),\mu_{\Gg,\al}(b'))
\ee
\be
s\nu_{\Gg,\al}(b,b')\eqdef\max(\nu_{\Gg,\al}(b),\nu_{\Gg,\al}(b'))
\ee
and
\be
i\nu_{\Gg,\al}(b,b')\eqdef\min(\nu_{\Gg,\al}(b),\nu_{\Gg,\al}(b'))\ .
\ee
We next define for any $q$, $1\le q\le p$, the expression
$\om(\Gg,{\bf h},q)$ as follows.

Let $l_q=\{b,b'\}$ for some $b\neq b'$ in $\cL$.

$\bullet$ If $b\in\Ga_{q-1,\Gg}$ and $b'\in W(\Ga_{q-1,\Gg})$, we
let
\be
\om(\Gg,{\bf h},q)\eqdef
\left\{
\begin{array}{ll}
0 & {\rm if}\ s\mu\ge i\nu\\
\frac{1}{h_{i\nu}}-\frac{1}{h_{s\mu}} & {\rm if}\ s\mu< i\nu
\end{array}
\right.
\ee
where $s\mu$ and $i\nu$ are shorthand for $s\mu_{\Gg,q-1}(b,b')$
and $i\nu_{\Gg,q-1}(b,b')$ respectively.
Note that $s\mu_{\Gg,q-1}(b,b')=s\mu_{\Gg}(b,b')\le q-1$
and  $i\nu_{\Gg,q-1}(b,b')=i\nu_{\Gg}(b,b')\le q-1$.

$\bullet$ If $b$, $b'\in W(\Ga_{q-1,\Gg})$, then we let
\be
\om(\Gg,{\bf h},q)\eqdef
-\frac{1}{h_{s\mu}}
\ee
where, again, $s\mu$ is shorthand for $s\mu_{\Gg,q-1}(b,b')$.
Note again that $s\mu_{\Gg,q-1}(b,b')=s\mu_{\Gg}(b,b')\le q-1$.

$\bullet$ Finally, in every other case for $b$ and $b'$, we let
\be
\om(\Gg,{\bf h},q)\eqdef 0\ .
\ee

Now let $l=\{b,b'\}$ be an unordered pair of elements of $\cL$,
such that $b=(\De,k)$ and   $b'=(\De',k')$; we then introduce the
functional derivation operator:
\be
D_l\eqdef
\int_\De dx \int_{\De'}dx'\ C(x,x')
\frac{\de}{\de\Ph(x,k)}
\frac{\de}{\de\Ph(x',k')}
\ee
We also introduce
\bea
\lefteqn{
\cR(\Gg,{\bf h})\eqdef
\int
d\mu_{\cC[\cM_{\Gg,{\bf h}}]}(\Ph)
\prod_{q=1}^p
\lp
\om(\Gg,{\bf h},q) D_{l_q}
\rp} & & \nonumber \\
 & & \prod_{i=1}^n
\Ph(x_i,0)
\exp
\lp
-\la\sum_{(\De,k)\in\cL_{\La,N}}
\int_\De
P(\Ph(x,k))dx
\rp
\eea
the functional derivations acting on any factor to their right.
We are now ready to state the main lemma for our expansion scheme.
\begin{lemma}
For any $m\ge 1$,
\bea
\lefteqn{
H_{\La,N}(x_1,\ldots,x_n)=} & & \nonumber \\
 & & \sum_{0\le p < m}
\sum_{\Gg=(l_1,\ldots,l_p)}
\int_{1>h_1>\cdots>h_p>0} dh_1\ldots dh_p
\ \cR(\Gg,(h_1,\ldots,h_p,0)) \nonumber \\
& & +\sum_{\Gg=(l_1,\ldots,l_m)}
\int_{1>h_1>\cdots>h_m>0} dh_1\ldots dh_m
\ \cR(\Gg,(h_1,\ldots,h_m,h_m))\ .
\label{maininter}
\eea
The sums on $\Gg$ are on all cluster-graphs with the prescribed
length.
\end{lemma}
\noindent{\bf Proof :}
We first prove the lemma for $m=1$.
For that we notice, according to equation (\ref{inter1}),
that
\be
H_{\La,N}(x_1,\ldots,x_n)=\cR(\Gg,{\bf h})
\ee
where $\Gg=\emptyset$ is the empty graph and ${\bf h}=(h_1)$
with $h_1=1$.
We then simply write
\be
H_{\La,N}(x_1,\ldots,x_n)=
\cR(\emptyset,(0))+\int_0^1 dh_1
\frac{d}{dh_1}\cR(\emptyset,(h_1))\ \ .
\ee
The covariance matrix appearing in $\cR(\emptyset,(h_1))$ is
\be
\cM_{\emptyset, (h_1)}=h_1\cM_\emptyset+(1-h_1)
T_{\Ga_0}[\cM_\emptyset]\ \ .
\ee

Therefore, the derivation with respect to $h_1$, produces a
functional derivation operator acting on the integrand, associated
to a matrix element of $\cM_\emptyset- T_{\Ga_0}[\cM_\emptyset]$
(this is obvious by Wick's theorem for polynomial integrands, then true 
for our smooth decreasing integrand by an easy limiting argument, see [8]).
That is we get a sum over $l_1=\{b,b'\}$ and a factor
$(\cM_\emptyset- T_{\Ga_0}[\cM_\emptyset])(b,b')D_{l_1}$
in the functional integral defining $\cR(\emptyset,(h_1))$.
It is a simple check to verify, with our previous definitions,
that
\bea
\lefteqn{
(\cM_\emptyset- T_{\Ga_0}[\cM_\emptyset])(b,b') } &  & \nonumber \\
 & & =\om(l_1,(h_1,h_1),1)\\
 & & =\left\{
\begin{array}{ll}
1 & {\rm if}\ b\in\Ga_0, b'\in W(\Ga_0)\cap\cL_0\\
-1 & {\rm if}\ b\neq b'\in W(\Ga_0)\ {\rm and}
\ \{b,b'\}\not\subset\cL_0\ .
\end{array}
\right.
\eea
Besides, the covariance matrix $\cM_{\emptyset,(h_1)}$ involved in
the functional integral can be rewritten, according to
(\ref{deccov}), as $\cM_{(l_1),(h_1,h_1)}$. Therefore
\be
H_{\La,N}(x_1,\ldots,x_n)=
\cR(\emptyset,(0))+\sum_{l_1}
\int_0^1 dh_1
\ \cR((l_1),(h_1,h_1))
\ee
which is the wanted result for $m=1$.

We now prove the induction step from $m\ge 1$ to $m+1$.
For this, we simply have to show that, given a cluster-graph
$\Gg=(l_1,\ldots,l_m)$ of length $m$ and parameters
$1>h_1>\cdots>h_m>0$, 
\bea
\lefteqn{
\cR(\Gg,(h_1,\ldots,h_m,h_m))=
\cR(\Gg,(h_1,\ldots,h_m,0))} & & \nonumber \\
 & & +\sum_{l_{m+1}}
\int_0^{h_m} dh_{m+1}
\ \cR((\Gg,l_{m+1}),(h_1,\ldots,h_m,h_{m+1},h_{m+1}))
\label{induc}
\eea
which is proven in the same way as for the $m=1$ case. Indeed, we write
\bea
\lefteqn{
\hskip -1.2cm\cR(\Gg,(h_1,\ldots,h_m,h_m))=} & & \nonumber\\
 & & \hskip -1.5cm \cR(\Gg,(h_1,\ldots,h_m,0))
\int_0^{h_m} dh_{m+1}
\frac{d}{dh_{m+1}}
\cR(\Gg,(h_1,\ldots,h_m,h_{m+1}))
\eea
and use (\ref{deccov}) to explicit the dependence on $h_{m+1}$ of
the covariance matrix:
\be
\cM_{\Gg,(h_1,\ldots,h_{m+1})}=
\frac{h_{m+1}}{h_m}\cM_{{\Gg'},(h_1,\ldots,h_{m})}
+\lp 1- \frac{h_{m+1}}{h_m}\rp
T_{\Ga_{m,\Gg}}[\cM_{\Gg',(h_1,\ldots,h_m)}]
\ee
where $\Gg'=(l_1,\ldots,l_{m-1})$.
Derivation with respect to $h_{m+1}$ again introduces a sum over
a new link $l_{m+1}=\{b,b'\}$, with a corresponding functional
derivation operator $D_{l_{m+1}}$ times a factor
\be
\frac{1}{h_m}\lp
\cM_{\Gg',(h_1,\ldots,h_m)}-
T_{\Ga_{m,\Gg}}[\cM_{\Gg',(h_1,\ldots,h_m)}]
\rp
(b,b')
\label{matel}
\ee
which is easily checked to be equal to
\be
\om((\Gg,l_{m+1}),(h_1,\ldots,h_{m+1},h_{m+1}),m+1)\ \ .
\ee
Indeed, if $b\neq b'\in W(\Ga_{m,\Gg})$, (\ref{matel}) is equal to
\be
\frac{1}{h_m}\lp
\cM_{\Gg',(h_1,\ldots,h_m)}(b,b')-1
\rp=\frac{1}{h_m}\lp
h_m\lp\frac{1}{h_m}-\frac{1}{h_{s\mu_{\Gg'}(b,b')}}\rp-1\rp
\ee
since $\nu_{\Gg'}(b)=\nu_{\Gg'}(b')=m$.
The situation $b\in\Ga_{m,\Gg}$, $b'\in W(\Ga_{m,\Gg})$
can be checked in the same way.

Finally the involved covariance matrix can be rewritten, thanks to
(\ref{deccov}), as
\be
\cM_{\Gg,(h_1,\ldots,h_{m+1})}=
\cM_{(\Gg,l_{m+1}),(h_1,\ldots,h_{m+1},h_{m+1})}
\ee
which proves (\ref{induc}).
\endproof

The easy proof that the cluster-graphs that are summed over in
lemma 3 satisfy the conditions (i) and (ii) stated earlier,
is left to the reader. We are now ready to move on to the proof of
theorem 1.

We first notice that, if $\Gg=(l_1,\ldots,l_p)$ is cluster-graph,
then $\# (\Ga_{p,\Gg})\ge p$; besides, the contribution of $\Gg$
in (\ref{maininter}) vanishes if $\Ga_{p,\Gg}$ is not contained in
$\cD_{\La,N}$ since a functional derivation $\frac{\de}{\de\Ph(x,k)}$
would have nothing to contract to. As a result,
$p>\#(\cD_{\La,N})$ implies that $\Gg=(l_1,\ldots,l_p)$ gives a
zero contribution; it is then straight-forward to take the limit
$m\rightarrow+\infty$ in (\ref{maininter}) to write
\bea
\lefteqn{
H_{\La,N}(x_1,\ldots,x_n)=} & & \nonumber\\
 & & \sum_{p=0}^{+\infty}
\sum_{{\Gg=(l_1,\ldots,l_p)}\atop{\Ga_{p,\Gg}\subset\cD_{\La,N}}}
\int_{1>h_1>\cdots>h_p>0} dh_1\ldots dh_p
\ \cR(\Gg,(h_1,\ldots,h_p,0))\ .
\label{maininter2}
\eea
We can now write an expression for the normalized Schwinger
functions since:
\bea
S_\La(x_1,\ldots,x_n) & = & \frac{H_{\La,N}(x_1,\ldots,x_n)}
{Z(\La)\times Z_0^{N|\La|}}\\
 & = &  \sum_{p=0}^{+\infty}
\sum_{{\Gg=(l_1,\ldots,l_p)}\atop{\Ga_{p,\Gg}\subset\cD_{\La,N}}}
\cA(\Gg,\La,N)
\label{expan}
\eea
where
\be
\cA(\Gg,\La,N)=
\frac{\cA_0(\Gg)}{Z_0^{\#(\Ga_{p,\Gg})}}\times
\frac{Z(Y_\Gg)\cdot Z_0^{\#(\La)-\#(Y_\Gg)}}{Z(\La)}
\label{factomain}
\ee
with the following notations.

$\bullet$ First, $Y_\Gg\eqdef\{\De\in\La|
h_{\Ga_{p,\Gg}}(\De)\ge N\}$.

$\bullet$ Next, $Z(Y_\Gg)$ is defined as in (\ref{zlambda}) by
\be
Z(Y_\Gg)\eqdef
\int d\mu_C(\Ph)
\exp\lp
-\la\sum_{\De\in Y_\Gg}\int_\De
P(\ph(x))dx
\rp
\ee
with a free boundary condition covariance.

$\bullet$ Finally, $\cA_0(\Gg)$ is defined, \und{independently of
$\La$ and $N$}, by
\bea
\lefteqn{
\cA_0(\Gg)=\int_{1>h_1>\cdots>h_p>0} dh_1\ldots dh_p
\int d\mu_{\cC[{\Br \cM}_{\Gg,(h_1,\ldots,h_p)}]}(\Ph)}
 & & \nonumber \\
 & & \prod_{q=1}^p \lp \om(\Gg,({\bf h},0),q)D_{l_q}
\rp \nonumber \\
 & & \prod_{i=1}^n \Ph(x_i,0)
\exp\lp
-\la\sum_{(\De,k)\in \Ga_{p,\Gg}}\int_\De
P(\Ph(x,k))dx
\rp
\eea
where
\be
{\Br\cM}_{\Gg,(h_1,\ldots,h_p)}(b,b')
\eqdef
\left\{
\begin{array}{ll}
\cM_{\Gg,(h_1,\ldots,h_p,0)}(b,b') & {\rm if}
\ b,b'\in\Ga_{p,\Gg}\\
0 & {\rm otherwise}.
\end{array}
\right.
\ee
The factorization (\ref{factomain}) stems from the fact that the
parameter vectors involved in (\ref{maininter2}) have a null last
component, and therefore the corresponding covariance matrix is
\be
\cM_{\Gg,(h_1,\ldots,h_p,0)}=
T_{\Ga_{p,\Gg}}[\cM_{\Gg,(h_1,\ldots,h_p,0)}]
\ee
which completely couples together the cubes of $W(\Ga_{p,\Gg})$
and decouples them from the rest of $\cL$.
This accounts for the factor $Z(Y_\Gg)$ which might be different
from $Z(\La)$, in case $\Ga_{p,\Gg}$ reaches the highest cubes of
$\cL_{\La,N}$ which contain all interaction terms of the form
$\exp(-\la\int_\De P(\Ph(x,k))dx)$.
For a given $\Gg$, $\cA(\Gg,\La,N)=\frac{\cA_0(\Gg)}
{Z_0^{\#(\Ga_{p,\Gg})}}$ as soon as
$N>\max\{h_{\Ga_{p,\Gg}}(\De)|\de\in\cD\}$ which is finite.
Besides, the only dependence in $\La$ is embodied in the condition
$\Ga_{p,\Gg}\subset \cL_{\La,N}$.

We will then show in the next section that there exists a positive
function $\cB(\Gg)$ of cluster-graphs $\Gg$, depending on $\la$,
such that, for small $\la$,
\be
\sum_\Gg \cB(\Gg)<+\infty
\ee
where the sum is without restriction on $\Gg$, and such that
\be
|\cA(\Gg,\La,N)|\le \cB(\Gg)
\ee
for any $\Gg$, $\La$, and $N$ satisfying $\Ga_{p,\Gg}\subset
\cL_{\La,N}$ and $N\ge \#(\La)$.

The discrete version of the Lebesgue dominated convergence theorem
will thus allow us to first take the limit $N\rightarrow +\infty$
and then the limit $\La\nearrow\RR^d$ in (\ref{expan}) thereby
proving theorem 1.
The next section is devoted to finding a uniform estimate $\cB(\Gg)$
which does the job.

\section{The uniform estimates}
\resetequ

We first use a very coarse bound for the ``parasite'' factors in
(\ref{factomain}).
\begin{lemma}
\be
0< \frac{1}{Z_0^{\#(\Ga_{p,\Gg})}}\times
\frac{Z(Y_\Gg)\cdot Z_0^{\#(\La)-\#(Y_\Gg)}}{Z(\La)}
\le \exp\lp
2 K_3 \la\#(\Ga_{p,\Gg})
\rp
\ee
where
\be
K_3\eqdef
K_2 \lp
1+\frac{(2m)!}{2^m m!}C(0,0)
\rp\ \ .
\ee
\end{lemma}

\noindent{\bf Proof :}
Indeed as we derived in section 3 a lower bound for $Z(\La)$,
it is easy to do the same with $Z_0$ and $Z(Y_\Gg)$, from which we
obtain the three estimates
\be
1\ge Z(\La)\ge \exp(-K_3\la\#(\La))
\ee
\be
1\ge Z(Y_\Gg)\ge \exp(-K_3\la\#(Y_\Gg))
\ee
and
\be
1\ge Z_0\ge \exp(-K_3\la)\ \ .
\ee
Now given $\Gg$, $\La$ and $N$, with $N\ge\#(\La)$, we have two
possible situations:

\noindent{\bf 1st case:} $Y_\Gg=\La$.

Then
\be
\frac{1}{Z_0^{\#(\Ga_{p,\Gg})}}\times
\frac{Z(Y_\Gg)\cdot Z_0^{\#(\La)-\#(Y_\Gg)}}{Z(\La)}
=Z_0^{-\#(\Ga_{p,\Gg})}
\le \exp
\lp
K_3 \la\#(\Ga_{p,\Gg})
\rp\ .
\ee

\noindent{\bf 2nd case:} $Y_\Gg\subset\La$ and $Y_\Gg\neq \La$.

Then
$N\le\max\{h_{\Ga_{p,\Gg}}(\De)|\de\in\cD\}$ from the remarks at
the end of section 3.
But $\#(\La)\le N$ and  $\max\{h_{\Ga_{p,\Gg}}(\De)|\de\in\cD\}\le
\#(\Ga_{p,\Gg})$
so that $\#(\La)\le \#(\Ga_{p,\Gg})$ and thus
\bea
\frac{1}{Z_0^{\#(\Ga_{p,\Gg})}}\times
\frac{Z(Y_\Gg)\cdot Z_0^{\#(\La)-\#(Y_\Gg)}}{Z(\La)}
 & \le & Z_0^{-\#(\Ga_{p,\Gg})}\cdot Z(\La)^{-1} \\
 & \le & \exp\lp 2 K_3 \la\#(\Ga_{p,\Gg})\rp\ .
\eea
\endproof

We now need a few lemmas to bound $\cA_0(\Gg)$.
\begin{lemma}
If $b=(\De,k)\in\Ga_{p,\Gg}$, and $\De'\in\cD$, then
\be
\sum_{k'\ge 0}
{\Br \cM}_{\Gg,(h_1,\ldots,h_p)}(b,(\De',k'))\le 1
\ee
\end{lemma}

\noindent{\bf Proof :}
Let us denote $b'=(\De',k')$.
Now only $b'\in\Ga_{p,\Gg}$ contributes.
Besides, either $b=b'$ or $s\mu_\Gg(b,b')<
i\nu_\Gg(b,b')$ is needed for
${\Br \cM}_{\Gg,(h_1,\ldots,h_p)}(b,b')\neq 0$.
Now remark that, for any $c\in \Ga_{p,\Gg}$,
$\mu_\Gg(c)\le j<\nu_\Gg(c)$ is equivalent to $c\in
W(\Ga_{i,\Gg})$. Therefore $s\mu_\Gg(b,b')<
i\nu_\Gg(b,b')$ means that there is $i$, $-1\le i\le p$,
such that both $b$ and $b'$ belong to $W(\Ga_{i,\Gg})$.

\noindent{\bf 1st case:} $\De=\De'$.

Since $W(\Ga)$ has a unique cube with a given $\De$, whatever is
the cluster $\Ga$, the only contribution comes from $k'=k$ which gives
1 and satisfies the inequality.

\noindent{\bf 2nd case:} $\De\neq\De'$.

Let $[k'_1,k'_2]\eqdef
\{k'|\exists i, \mu_\Gg(b)\le i<\nu_\Gg(b), (\De,k')\in
W(\Ga_{i,\Gg})
\}$.
Let us first suppose that $k'_2\ge k'_1+1$.
We let $\mu_{k'}\eqdef\mu_\Gg((\De',k'))$
and $\nu_{k'}\eqdef\nu_\Gg((\De',k'))$.
If $b'=(\De',k')$ with $k'_1<k'<k'_2$, it follows from the
definition of a cluster-graph like $\Gg$ that we have
$\nu_{k'}=\mu_{k'+1}$,
$\mu_{k'}>\mu_{\Gg}(b)$ and $\nu_{k'}<\nu_{\Gg}(b)$.
Therefore
\bea
\hskip-1cm\sum_{k'_1<k'<k'_2}
{\Br \cM}_{\Gg,(h_1,\ldots,h_p)}(b,(\De',k')) & = &
\sum_{k'_1<k'<k'_2} h_{\nu_\Gg(b)}
\lp
\frac{1}{h_{\mu_{k'+1}}}-\frac{1}{h_{\mu_{k'}}}
\rp \\
 & = &  h_{\nu_\Gg(b)}
\lp
\frac{1}{h_{\mu_{k'_2}}}-\frac{1}{h_{\mu_{k'_1+1}}}
\rp\ \ .
\eea
One also checks easily that the contribution of $k'=k'_1$ is
\be
h_{\nu_\Gg(b)}
\lp
\frac{1}{h_{\mu_{k'_1+1}}}-\frac{1}{h_{\mu_{\Gg}(b)}}
\rp
\ee
and that of $k'=k'_2$ is
\be
h_{\nu_{k'_2}}
\lp
\frac{1}{h_{\nu_{\Gg}(b)}}-\frac{1}{h_{\mu_{k'_2}}}
\rp\ \ .
\ee
Therefore
\be
\sum_{k'\ge 0}
{\Br \cM}_{\Gg,(h_1,\ldots,h_p)}(b,(\De',k'))
=
\frac{h_{\nu_{\Gg}(b)}}{h_{\mu_{k'_2}}}-
\frac{h_{\nu_{\Gg}(b)}}{h_{\mu_{\Gg}(b)}}
+\frac{h_{\nu_{k'_2}}}{h_{\nu_{\Gg}(b)}}-
\frac{h_{\nu_{k'_2}}}{h_{\mu_{k'_2}}}\ \ .
\ee
But, from $\mu_\Gg(b)<\mu_{k'_2}<\nu_\Gg(b)\le\nu_{k'_2}$,
it follows that there is $\al$, $\beta$, $\ga\in [0,1]$ such that
$h_{\nu_{k'_2}}=\al h_{\nu_{\Gg}}(b)$,
$h_{\nu_{\Gg}}(b)=\beta h_{\mu_{k'_2}}$ and
$h_{\mu_{k'_2}}=\ga h_{\mu_{k'_2}}$.
Thus
\bea
\sum_{k'\ge 0}
{\Br \cM}_{\Gg,(h_1,\ldots,h_p)}(b,(\De',k'))
& = &
\beta-\beta\ga+\al-\al\beta\\
 & \le &  \beta+\al-\al\beta\\
 & \le &  1-(1-\al)(1-\beta)\\
 & \le & 1
\eea
which proves the assertion.
\endproof

As a consequence of this lemma we have a bound
\be
\left|
\cC[{\Br \cM}_{\Gg,(h_1,\ldots,h_p)}]
(x,k;x',k')
\right|
\le
G((\De(x),k);(\De(x'),k'))
\ee
where the function $G(b,b')$ on $\cL^2$ satisfies
\be
\forall b\in\cL,\;\;
\sum_{b'\in\cL}
G(b,b')\le K_4
\ee
for some constant $K_4$. Indeed,
\bea
\lefteqn{
G((\De,k);(\De',k'))\eqdef} & & \nonumber \\
 & & \hskip -1cm{\Br \cM}_{\Gg,(h_1,\ldots,h_p)}
((\De,k);(\De',k'))
\times K_1(d+1)
\times (1+d(\De,\De'))^{-(d+1)}
\eea
with $d(\De,\De')\eqdef\min\{|x-y|\ |\ x\in\De,y\in\De'\}$
works, since the sum over $k'$, by lemma 5,
is no greater than 1, and the sum over
$\De'$ is bounded by the rapid decay (\ref{rapdec}) of the
propagator. Note that $K_4$, unlike $G(b,b')$, is independent of
$\Gg$ and $(h_1,\ldots,h_p)$.
\begin{lemma}
(The principle of local factorials)

We have the bound:
\be
\left|
\int d\mu_{\cC[{\Br \cM}_{\Gg,(h_1,\ldots,h_p)}]}(\Ph)
\ \Ph(z_1,k_1)\cdots
\Ph(z_r,k_r)
\right|
\le K_5^5\times\prod_{b\in\cL}\sqrt{n(b)!}
\ee
where $n(b)\eqdef
\#(\{j| 1\le j\le r, (\De(z_j),k_j)=b_j\})$ and $K_5$ is a constant.
\end{lemma}

\noindent{\bf Proof :}
Using Wick's theorem, the functional integral can be computed as a
sum over contractions $c$ of the fields $\Ph(z_j,k_j)$, with the
propagator of $\cC[{\Br \cM}_{\Gg,(h_1,\ldots,h_p)}]$. $c$ is
simply an involution without fixed points of the set
$J=\{1,\ldots,r\}$. We get
\bea
\lefteqn{
\left|
\int d\mu_{\cC[{\Br \cM}_{\Gg,(h_1,\ldots,h_p)}]}(\Ph)
\ \Ph(z_1,k_1)\cdots
\Ph(z_r,k_r)
\right|} &  &  \nonumber\\
 & & =\left| \sum_{c}\prod_{{\{j,j'\}\subset J}\atop{j'=c(j)}}
\cC[{\Br \cM}_{\Gg,(h_1,\ldots,h_p)}]
(x,j;x_{c(j)},k_{c(j)})
\right| \\
 & & \le  \sum_{c}\prod_{{\{j,j'\}\subset J}\atop{j'=c(j)}}
 G(b_j,b_{c(j)})
 \eea
where $b_j$ denotes $(\De(x_j),k_j)\in\cL$.
Suppose we have ordered $J$ as $\{j_1,\ldots,j_s\}$
such that $n(b_{j_1})\ge n(b_{j_2})\ge\cdots\ge n(b_{j_s})$.
To sum over $c(j_1)$, we first sum over $b_{c(j_1)}$, then
over $c(j_1)$ knowing  $b_{c(j_1)}$.
The sum over $b_{c(j_1)}$ is bounded by $K_4$.
The sum over  $c(j_1)$ knowing  $b_{c(j_1)}$ costs a factor
$n(b_{c(j_1)})\le\sqrt{n(b_{j_1})n(b_{c(j_1)})}$ because of the
ordering of $J$. We now pick the element $j$ with the smallest
label in $J\backslash\{j_1,c(j_1)\}$, and sum over $c(j)$ in the
same way, thus getting a factor $K_4\sqrt{n(b_{j})n(b_{c(j)})}$,
and so on.
Since $\sqrt{n(b_j)}$ will appear exactly once by definition of a
contraction $c$, we obtain a bound
\bea
K_4^{\frac{r}{2}}\times\prod_{j\in J}\sqrt{n(b_j)} & = &
K_4^{\frac{r}{2}}\times\prod_{{b\in\cL}\atop{n(b)\neq 0}}
\sqrt{n(b)^{n(b)}}\\
 & \le & K_5^r \prod_{b\in\cL}\sqrt{n(b)!}
\eea
with $K_5\eqdef \sqrt{e K_4}$.
\endproof

We now explain the bound on $\cA_0(\Gg)$. First note that $\cA_0(\Gg)$
decomposes as
\be
\cA_0(\Gg)=\sum_\rh \cA_0(\Gg,\rh)
\ee
where $\rh$ is a derivation procedure for the operators $D_{l_q}$
and  $\cA_0(\Gg,\rh)$ is the contribution of $\rh$ in the
expansion that computes the action of $\prod_{q=1}^p D_{l_q}$ on
the integrand
\be
\prod_{i=1}^n\Ph(x_i,0)
\exp\lp
-\la\sum_{(\De,k)\in\Ga_{p,\Gg}}\int_\De
P(\Ph(x,k))dx\rp\ \ .
\ee
When considering the expression for $\cA_0(\Gg,\rh)$,
we take out of the functional integral all the $\om(\Gg,({\bf h},0),q)$
factors, as well as the $C(x,x')$ factors coming from
$\prod_{q=1}^p D_{l_q}$, and also the spatial integrations
$\int_\De dx$ that come from the $D_{l_q}$, as well as all
numerical factors such as $\la$ or the coefficients of the polynomial
$P$.

The resulting expression is a functional integral of the form:
\bea
\lefteqn{
\cI=
\int d\mu_{\til \cC}(\Ph)
\ \Ph(z_1,k_1)\ldots\Ph(z_r,k_r)} & & \nonumber \\ 
 & & \exp\lp
-\la\sum_{(\De,k)\in\Ga_{p,\Gg}}\int_\De
P(\Ph(x,k))dx\rp
\eea
where ${\til \cC}$ denotes $\cC[{\Br\cM}_{\Gg,(h_1,\ldots,h_p)}]$.
We bound it using
\be
|\cI|\le
\int d\mu_{\til \cC}(\Ph)
\ |\Ph(z_1,k_1)\ldots\Ph(z_r,k_r)|
\exp(\la K_6\#(\Ga_{p,\Gg}))
\ee
where $K_6=\min\{P(x)|x\in\RR\}$. Then by the Cauchy-Schwartz
inequality,
\be
|\cI|\le
\exp(\la K_6\#(\Ga_{p,\Gg}))
\sqrt{
\int d\mu_{\til \cC}(\Ph)
\ \Ph(z_1,k_1)^2\ldots\Ph(z_r,k_r)^2}\ .
\ee
Now we bound the functional integral in the last inequality using
lemma 6 thus obtaining:
\be
|\cI|\le
\exp(\la K_6\#(\Ga_{p,\Gg}))\times
K_5^r\times
\prod_{b\in\cL}
(2n_{\Gg,\rh}(b))!^{\frac{1}{4}}
\label{boundi}
\ee
where
$n_{\Gg,\rh}(b)\eqdef\#(\{j|1\le j\le r,(\De(z_j),k_j)=b\})$.

We now explain the bound on the sum over the derivation procedures
$\rh$ that act on
\be
\prod_{i=1}^n\Ph(x_i,0)
\exp\lp
-\la\sum_{(\De,k)\in\Ga_{p,\Gg}}\int_\De
P(\Ph(x,k))dx\rp\ \ .
\ee

First we bound the propagators $C(x,x')$ corresponding to a
$D_{l_q}$ with $l_q=\{(\De,k),(\De',k')\}$
by $K_1(r)(1+d(\De,\De'))^{-r}$. The exponent $r$ will be adjusted later.
We also bound the spatial integrations $\int_\De dx$ by 1.
Since each $(\De,k)\in\Ga_{p,\Gg}\backslash\Ga_0$ belongs to an
$l_q$, there is at least a $\frac{\de}{\de\Ph}$ that acts on the
corresponding interaction term $\exp(-\la\int_\De
P(\Ph(x,k))dx)$; therefore there is at least
$\la^{\#(\Ga_{p,\Gg})-\#(\Ga_0)}$ in factor and eventually some
more factors $\la$ that we bound by 1 as we assume from now on
that $\la\le 1$.

We also introduce the notation $||P||$ for the maximum of absolute
value of the coefficients of the polynomial $P$.
Note that each $\frac{\de}{\de\Ph(x,k)}$ can derive an interaction
term, and thus generate a coefficient of $P$. We therefore
globally bound these factors by $(1+||P||)^{2p}$.
We let $n_\Gg\eqdef\#(\{q|1\le q\le p, b\in l_q\})$,
i.e. the coordinance of $b$ with respect to the graph $\Gg$, for
any $b\in\Ga_{p,\Gg}$.
We also let $s(b)\eqdef\#(\{i| 1\le i\le n, b=(\De(x_i),0)\})$
that counts the sources located in $b$.

Choose an arbitrary order to perform the functional derivations.
Let $\frac{\de}{\de(x,k)}$ be the one performed last.
It is located in $b=(\De(x),k)$, and can either derive one of the
sources, which gives $s(b)$ possibilities. It can also derive a
new vertex from the interaction $\exp(-\la\int_{\De(x)}
P(\Ph(y,k))dy)$, we then have to choose the derived monomial in
$P$, and the field in the monomial which gives at most $(2m)^2$
new possibilities.
Finally it can rederive a vertex that was derived for the first
time by a previously performed functional derivation
$\frac{\de}{\de\Ph(x',k)}$ that is also located in $b$.
This gives a total number of possibilities, for $\frac{\de}
{\de\Ph(x,k)}$, that is bounded by $s(b)+4m^2n_\Gg(b)$.

We then do the same sum over the ways of computing the before
last functional derivation, and so on.
It follows that the number of derivation procedures $\rh$ is
bounded by
\be
\prod_{b\in\Ga_{p,\Gg}}\lp
s(b)+4m^2n_\Gg(b)
\rp^{n_\Gg(b)}
\ee
since there is $n_\Gg(b)$ functional derivations in each $b$.
We write for convenience
\bea
\prod_{b\in\Ga_{p,\Gg}}\lp
s(b)+4m^2n_\Gg(b)
\rp^{n_\Gg(b)} & \le &
\prod_{b\in\Ga_{p,\Gg}}\lp
n_\Gg(b)!
e^{s(b)+4m^2n_\Gg(b)}\rp \\
 & \le & e^{n+8 m^2 p} \prod_{b\in\Ga_{p,\Gg}}
n_\Gg(b)!\ \ .
\eea
Now note that in (\ref{boundi}), $r\le n+4 m p$, and
for each $b$, $n_{\Gg,\rh}\le s(b)+ 2 m n_\Gg(b)$.
As a result, the previous bound on $\cI$ becomes
\bea
\hskip -1cm |\cI| & \le &\hskip -.3cm
\exp(K_6\#(\Ga_{p,\Gg}))\times
(1+K_5)^{n+4mp}
\prod_{b\in\Ga_{p,\Gg}}\lp 2s(b)+4m n_\Gg(b)
\rp!^{\frac{1}{4}} \\
 & \le & \exp(K_6\#(\Ga_{p,\Gg}))\times
(1+K_5)^{n+4mp} \nonumber \\
 &  &  \times\prod_{b\in\Ga_{p,\Gg}}\lp
 \sqrt{s(b)!}\times (n_\Gg(b)!)^m\times
 \exp(3m s(b)+6m^2 n_\Gg(b))\rp \\
 & \le & \exp(K_6\#(\Ga_{p,\Gg}))\times
(1+K_5)^{n+4mp} \nonumber \\
 & & \times\sqrt{n!}\times
e^{3mn+12m^2p}\times \prod_{b\in\Ga_{p,\Gg}}
(n_\Gg(b)!)^m\ \ .
\eea
We are now able to write a raw bound on $\cA_0(\Gg)$ as:
\bea
|\cA_0(\Gg)| & \le & \la^{\#(\Ga_{p,\Gg})-\#(\Ga_0)}\times
\prod_{q=1}^p\lp
K_1(r)(1+d(\De_q,\De'_q))^{-r}\rp \nonumber \\
 & & \times \int_{1>h_1>\cdots>h_p>0} dh_1\ldots dh_p
\prod_{q=1}^p  |\om(\Gg,({\bf h},0),q)| \nonumber \\
 & & \times (1+||P||)^{2p}\times \exp(K_6\#(\Ga_{p,\Gg}))\times
(1+K_5)^{n+4mp} \nonumber \\
 & & \times \sqrt{n!} \times e^{(3m+1)n+20m^2 p}
\times  \prod_{b\in\Ga_{p,\Gg}}
(n_\Gg(b)!)^{m+1}
\label{boundazero}
\eea
where $\De_q$, $\De'_q$ are such that $l_q=\{(\De_q,k_q),
(\De'_q,k'_q)\}$, for some $k_q$ and $k'_q$.

The right-hand side is not quite $\cB(\Gg)$, we need first to get
rid of the local factorials $n_\Gg(b)!$. This requires a volume
argument and the next two lemmas.
\begin{lemma}
If $\Gg=(l_1,\ldots,l_p)$ is a cluster-graph with
$\cA_0(\Gg)\neq 0$, and $l_{q_\al}=\{b_\al,b'_\al\}$,
$1\le \al\le 3$, are three links in $\Gg$ such that $q_1< q_2<
q_3$ and $b_1=b_2=b_3$; then $b'_1$, $b'_2$ and $b'_3$ cannot
all be of the form $(\De',k'_\al)$ with the same $\De'\in\cD$.
\end{lemma}

\noindent{\bf Proof :}
{\em Ad absurdum}. Let $b=b_1=b_2=b_3=(\De,k)$,
and $b'_\al=(\De',k'_\al)$, $1\le \al \le 3$.
Since $l_q\not\subset\Ga_{q-1,\Gg}$ for any $q$,
and since $q_1< q_2<
q_3$ we have that $k'_1$, $k'_2$ and $k'_3$ are distinct. We even
have  $k'_1<k'_2<k'_3$. Indeed, if for
instance $k'_2<k'_1$, since
$l_{q_1}=\{b,(\De',k'_1)\}\subset\Ga_{q_1,\Gg}$ and $\Ga_{q_1,\Gg}$
is a cluster, it would follow that $(\De',k'_2)\in\Ga_{q_1,\Gg}$
and thus $l_{q_2}\subset\Ga_{q_1,\Gg}\subset\Ga_{q_2-1,\Gg}$
which is not allowed.

Now if we only consider $l_{q_1}$ and  $l_{q_2}$, since
$b\in\Ga_{q_2-1,\Gg}$, $l_{q_2}$ can only be of type cluster-roof,
and $\om(\Gg,({\bf h},0),q_2)\neq 0$ implies
$s\mu_{\Gg,q_2-1}(b,b'_2)< i\nu_{\Gg,q_2-1}(b,b'_2)$.
That is, there exists $q< q_2$ such that $b$, $b'_2\in
W(\Ga_{q,\Gg})$. Thus $b\notin\Ga_{q,\Gg}$ and therefore $q<q_1$.
Besides, $b'_2\in W(\Ga _{q,\Gg})$ and $k'_2>k'_1$
implies $b'_1\in\Ga_{q,\Gg}\subset\Ga_{q_1-1,\Gg}$.
But $l_{q_1}\not\subset\Ga_{q_1-1,\Gg}$, therefore $b\notin
\Ga_{q_1-1,\Gg}$. As a result, $\mu_\Gg(b'_1)<\mu_\Gg(b)=q_1-1$.

We can now do the same reasoning, considering $l_{q_2}$ and
$l_{q_3}$ this time, to conclude $\mu_\Gg(b'_2)<\mu_\Gg(b)=q_2-1$
as well, which gives a different value for $\mu_\Gg(b)$ and
proves a contradiction.
\endproof

\begin{lemma}
(The volume argument)

We have, with the notations of (\ref{boundazero}),
\be
\prod_{b\in\Ga_{p,\Gg}}
(n_\Gg(b)!)^{m+1}
\times
\prod_{q=1}^p
(1+d(\De_q,\De'_q))^{-r_1}
\le
K_7^p
\ee
for some constants $r_1$ and $K_7$ that only depend on the
dimension $d$ and the degree $2m$ of the interaction.
\end{lemma}

\noindent{\bf Proof :}
We let $r_1= 4d(m+2)$. We now write
\be
\prod_{b\in\Ga_{p,\Gg}}
(n_\Gg(b)!)^{m+1}
\times
\prod_{q=1}^p
(1+d(\De_q,\De'_q))^{-r_1}
=
\prod_{{b\in\Ga_{p,\Gg}}\atop{n_\Gg(b)\ge 1}}
\xi(b)
\ee
with
\be
\xi(b)\eqdef
n_\Gg(b)!\times\prod_{b'\ {\rm linked\ to}\ b}
(1+d(\De(b),\De(b')))^{-\frac{r_1}{2}}
\ee
where the product is over all $b'\in\Ga_{p,\Gg}$ such that
$\{b,b'\}$ is a link of $\Gg$, and $\De(b)$ denotes the first
projection on $\cD$ of the pair $b\in\cL$.
Now it follows from lemma 7 that there cannot be more than two
cubes $b'$, with the same $\De(b')$, linked to $b$.
Remark that there is a constant $K$ such that for $\de$ big enough
\be
\#(\{\De'\in\cD|d(\De(b),\De')\le \de\})\le K \de^d\ \ .
\ee
Therefore
\be
\#(\{b'\in\Ga_{p,\Gg}|\ b'\ {\rm linked\ to}\ b,
d(\De(b),\De(b'))\le \de\})\le 2K \de^d\ \ .
\ee
If $n_\Gg(b)$ is big enough  and if we set
$\de=(\frac{n}{4K})^\frac{1}{d}$,
it follows that at least $\frac{n_\Gg(b)}{2}$ cubes $b'$ that are
linked to $b$ satisfy $d(\De(b),\De(b'))>\de$.
As a result:
\bea
\xi(b) & \le &  (n_\Gg(b)!)^{m+1}\times(1+\de)^{-\frac{r_1 n_\Gg(b)}
{4}} \\
 & \le & n_\Gg(b)^{(m+1)n_\Gg(b)}\times \lp
\frac{n_\Gg(b)}{4K}\rp^{-\frac{r_1 n_\Gg(b)}
{4d}} \\
 & \le & n_\Gg(b)^{-n_\Gg(b)}\times (4K)^{\frac{r_1 n_\Gg(b)}
{4d}}
\eea
because of our choice for $r_1$.
It easily follows that $\xi(b)\le K'$ for some constant $K'\ge 1$,
for any value of $n_\Gg(b)$.
Taking $K_7\eqdef {K'}^2$ concludes the proof of the lemma.
\endproof

We now return to (\ref{boundazero}) and proceed to define the
bounding term $\cB(\Gg)$. First we choose $r=r_1+d+1$.
Next we note that $\#(\Ga_{p,\Gg})-\#(\Ga_0)\ge p$ and
$\#(\Ga_{p,\Gg})\le 2p+n$.
Combining lemma 4, (\ref{boundazero}) and lemma 8,
we now easily obtain a bound
\bea
\lefteqn{
|\cA(\Gg,\La,N)|  \le  K_8(n) K_9^p \la^p 
\times  \int_{1>h_1>\cdots>h_p>0} dh_1\ldots dh_p} & & \nonumber \\
 & & \prod_{q=1}^p
\lp
|\om(\Gg,({\bf h},0),q)|(1+d(\De_q,\De'_q))^{-(d+1)}
\rp
\label{boundall}
\eea
where $K_8(n)$ and $K_9$ are independent of $\Gg$, $\La$
and $N$.
We let $\cB(\Gg)$ be the righthand side of (\ref{boundall}).
The proof of theorem 1 will be complete when we prove the
following result.
\begin{prop}
There exists $\la_0>0$ such that for any $\la\in[0,\la_0]$,
\be
\sum_\Gg \cB(\Gg) < +\infty
\ee
where the cluster-graph $\Gg$ is summed without any restriction of
volume in $\cL$.
\end{prop}

\noindent{\bf Proof :}
For any cluster-graph $\Gg$ with nonzero contribution, we define
the following function
$\si_\Gg:\{1,\ldots,p\}\rightarrow\{0,\ldots,p-1\}$.
Let $q$, $1\le q\le p$, and $l_q=\{b_q,b'_q\}$,
and let ${\Br b}_q$ and ${\Br b}'_q$ be the two elements of
$W(\Ga_{q-1,\Gg})$ with the same first projection on $\cD$
as $b_q$ and $b'_q$ respectively.
We pose, by definition,
\be
\si_\Gg(q)\eqdef\max
(\mu_\Gg({\Br b}_q),\mu_\Gg({\Br b}'_q))< q\ \ .
\ee
Note that, indeed, $\si_\Gg(q)\ge 0$, otherwise we would have
${\Br b}_q$, ${\Br b}'_q\in W_{-1}=\cL_0$ and therefore also
$b_q$, $b'_q\in W_{-1}$, which would give
$\om(\Gg,({\bf h},0),q)=0$ and a zero contribution for $\Gg$.
We will first bound the conditional sum on $\Gg$, knowing
$\si_\Gg$.

We start by summing over the last link $l_p$ knowing
$\Gg'=(l_1,\ldots,l_{p-1})$ and  $\si_\Gg$. We first perform the
sum over $l_p=\{b_p,b'_p\}$ with $b_p=(\De_p,k)$ and
$b'_p=(\De'_p,k')$, knowing $\De_p$ and $\De'_p$.
This is done thanks to the factor $|\om(\Gg,({\bf h},0),p)|$ as
in lemma 5.
Note that there are three cases.

\medskip
\noindent{\bf 1st case:} $l_p$ is a roof-roof link.

In this situation $b_p$, $b'_p\in W(\Ga_{p-1,\Gg})$ and thus
$b_p={\Br b}_p$, $b'_p={\Br b}'_p$ and
\be
|\om(\Gg,({\bf h},0),p)|=
\left|
-\frac{1}{h_{s\mu_\Gg(b_p,b'_p)}}
\right|
=\frac{1}{h_{\si_\Gg(p)}}\ \ .
\ee

\medskip
\noindent{\bf 2nd case:} $l_p$ is cluster-roof, with $b_p\in
W(\Ga_{p-1,\Gg})$ and $b'_p\in\Ga_{p-1,\Gg}$.

Then $b_p={\Br b}_p$ is unique, and we have to sum over the second
projection $k'$ of $b'_p$, $0\le k'\le
h_{\Ga_{p-1,\Gg}}(\De'_p)$, with the condition that
$s\mu_\Gg(b_p,b'_p)<i\nu_\Gg(b_p,b'_p)$.
We obtain, the previous condition being implicit in the following
sums,
\be
\sum_{k'}|\om(\Gg,({\bf h},0),p)|=
\sum_{k'}\lp
\frac{1}{h_{i\nu_\Gg(b_p,b'_p)}}-\frac{1}{h_{s\mu_\Gg(b_p,b'_p)}}
\rp\ \ .
\ee
Note that $\De_p\neq\De'_p$ as no link is vertical.
We let
\bea
\lefteqn{
[k'_1,k'_2]\eqdef\{k'|0\le k'\le
h_{\Ga_{p-1,\Gg}}(\De'_p)} & & \nonumber \\
& & {\rm and\ }
\exists i, \mu_\Gg(b_p)\le i\le p-1,
(\De'_p,k')\in W(\Ga_{i,\Gg})\}\ .
\eea
With the notation $\mu_{k'}=\mu_\Gg((\De'_p,k'))$ and
$\nu_{k'}=\nu_\Gg((\De'_p,k'))$, we have that for any $k'$,
$k'_1\le k'<k'_2$, $\mu_{k'+1}=\nu_{k'}$.
Note also that $\nu_{k'_2}=\mu_\Gg({\Br b}'_p)$.
Therefore
\bea
\lefteqn{
\sum_{k'}\lp
\frac{1}{h_{i\nu_\Gg(b_p,b'_p)}}-\frac{1}{h_{s\mu_\Gg(b_p,b'_p)}}
\rp =} & & \nonumber \\
 &  & \sum_{k'_1<k'<k'_2}\lp
\frac{1}{h_{\mu_{k'+1}}}-\frac{1}{h_{\mu_{k'}}}
\rp  +\lp\frac{1}{h_{\mu_{k'_1+1}}}-\frac{1}{h_{\mu_{\Gg}(b_p)}}
\rp \nonumber \\
 & & +\lp\frac{1}{h_{\mu_{\Gg}({\Br b}'_p)}}-\frac{1}{h_{\mu_{k'_2}}}
\rp \\
 &  & =\frac{1}{h_{\mu_{\Gg}({\Br b}'_p)}}-
\frac{1}{h_{\mu_{\Gg}(b_p)}}
\eea
which is positive; since $\mu_\Gg({\Br b}'_p)\ge\mu_\Gg(b_p)$ is
necessary for the existence of cluster-roof links $\{b_p,b'_p\}$
with $b'_p$ under ${\Br b}'_p$.
Finally
\be
\sum_{k'}|\om(\Gg,({\bf h},0),p)|\le
\frac{1}{h_{\mu_\Gg({\Br b}'_p)}}=\frac{1}{h_{\si_\Gg(p)}}\ \ .
\ee

\medskip
\noindent{\bf 3rd case:} $l_p$ is cluster-roof, with $b'_p\in
W(\Ga_{p-1,\Gg})$ and $b_p\in\Ga_{p-1,\Gg}$.

The symmetric of the 2nd case is treated in the same way, giving a
bound of $\frac{1}{h_{\si_\Gg(p)}}$ again.

So summing on $l_p$, knowing $\De_p$ and $\De'_p$, gives
a bound of $\frac{3}{h_{\si_\Gg(p)}}$.

We then need to sum over the unordered pair $\{\De_p,\De'_p\}$,
knowing $\Gg'=(l_1,\ldots,l_{p-1})$ and $\si_\Gg$.
Note that one of the cubes ${\Br b}_p$ and ${\Br b}'_p$
has a $\mu_\Gg$ equal to $\si_\Gg(p)$.
Assume it is ${\Br b}_p$ for instance. Since $\si_\Gg(p)=
\mu_\Gg({\Br b}_p)\ge 0$, we have that ${\Br b}_p\notin\cL_0$.
There is then a unique box $\und{b}$ just under ${\Br b}_p$,
i.e. such that $\und{b}=(\De_p,k-1)$ if ${\Br b}_p=(\De_p,k)$.
We then have $\nu_\Gg(\und{b})=\mu_\Gg({\Br b}_p)=\si_\Gg(p)$.

Either $\si_\Gg(p)=0$, in this case $\und{b}\in\Ga_{0}$, for which
there is at most $\#(\Ga_0)\le n$ possibilities.
Or $\si_\Gg(p)>0$; in that case $\und{b}\in l_{\si_\Gg(p)}
\backslash\Ga_{\si_\Gg(p)-1,\Gg}$ which leaves two possibilities.
Once we know $\und{b}$, we know one of the elements of $\{\De_p,
\De'_p\}$. The sum over the other one is done thanks to the factor
$(1+d(\De_p,\De'_p))^{-(d+1)}$, and is bounded by some constant.
As a result
\bea
\lefteqn{
\sum_{l_p}
|\om(\Gg,({\bf h},0),p)|(1+d(\De_p,\De'_p))^{-(d+1)}}
 & & \nonumber \\
 & & \le\frac{K_{10}}{h_{\si_\Gg(p)}}
\lp \bbbone_{\{\si_\Gg(p)>0\}}+n\bbbone_{\{\si_\Gg(p)=0\}}
\rp
\eea
for some constant $K_{10}$,
the sum being over $l_p$ knowing $(l_1,\ldots,l_{p-1})$ and
the full map $\si_\Gg$.
$\bbbone_{\{...\}}$ denotes the characteristic function of the
event between braces.

We can now repeat the operation and sum over $l_{p-1}$ knowing
$(l_1,\ldots,l_{p-2})$ and $\si_\Gg$; and so on.
We then get
\bea
\sum_{{\Gg\ {\rm of}}\atop{{\rm length\ }p}}
\cB(\Gg)
 & \le & \sum_{\si} K_8(n)K_9^p \la^p K_{10}^p
\times  \int_{1>h_1>\cdots>h_p>0} dh_1\ldots dh_p \nonumber \\
 & & \prod_{q=1}^p \frac{\bbbone_{\{\si(p)>0\}}
+n\bbbone_{\{\si(p)=0\}}}{h_{\si(p)}}
\eea
where the sum is over all maps $\si:\{1,\ldots,p\}\rightarrow
\{0,\ldots,p-1\}$ such that $\si(q)<q$ for any $q$, $1\le q\le p$.
The last step relies on the following lemma.
\begin{lemma}
For any $p\ge 1$, any $J=\{j_1,\ldots,j_\al\}\subset
\{1,\ldots,p\}$ with $j_1<\cdots<j_\al$, we have
\be
\sum_{\si|J} \int_{1>h_1>\cdots>h_p>0} dh_1\ldots dh_p
\ \prod_{q=1}^p \frac{1}{h_{\si(q)}}
\le \frac{e^p}{\al!}
\ee
where the sum is over maps $\si:\{1,\ldots,p\}\rightarrow
\{0,\ldots,p-1\}$ such that for any $q\in J$, $\si(q)=0$
and for any $q\notin J$, $1\le \si(q)<q$.
\end{lemma}

\noindent{\bf Proof of the lemma :}
We perform a change of variables by letting
$h_q=s_1 s_2\ldots s_q$, $1\le q\le p$, so that
\be
\int_{1>h_1>\cdots>h_p>0} dh_1\ldots dh_p
\prod_{q=1}^p \frac{1}{h_{\si(q)}}
=\int_0^1 ds_1\ldots\int_0^1 ds_p  \prod_{q=1}^p
\lp\prod_{\si(q)<j<q}s_j
\rp
\ee
and
\be
\sum_{\si|J} \int_{1>h_1>\cdots>h_p>0} dh_1\ldots dh_p
\prod_{q=1}^p \frac{1}{h_{\si(q)}}
= \int_0^1 ds_1\ldots\int_0^1 ds_p  \prod_{q=1}^p P_q(s)
\label{decalan}
\ee
where
\be
P_q(s)\eqdef
\left\{
\begin{array}{ll}
s_1 s_2\ldots s_{q-1} & {\rm if}\ q\in J\\
1+s_{q-1}+s_{q-1}s_{q-2}+\cdots+s_{q-1}s_{q-2}\ldots s_2 & {\rm if}
\ q\notin J\ .
\end{array}
\right.
\ee
Suppose $q\notin J$ and $q+1\in J$. The product of the
corresponding factors is then
\bea
\lefteqn{
(1+s_{q-1}+s_{q-1}s_{q-2}+\cdots+s_{q-1}s_{q-2}\ldots s_2)
s_1 s_2\ldots s_{q}
} & & \nonumber \\
 & &  \le (1+s_{q-1}+s_{q-1}s_{q-2}+\cdots+s_{q-1}s_{q-2}\ldots s_2)
s_1 s_2\ldots s_{q} \nonumber \\
 & & \ \  +  s_1 s_2\ldots s_{q-1} \\
 & & =s_1 s_2\ldots s_{q-1} (1+s_{q}+s_{q}s_{q-1}+\cdots
+s_{q}s_{q-1}\ldots s_2)
\eea
which is the product we would get if the opposite situation
occurred that is $q\in J$ and $q+1\notin J$.
Therefore, if we lower the elements of $J$, one by one, in
$\{1,\ldots,p\}$ we maximize the righthand side of
(\ref{decalan}),
and we only need to prove the bound for
\bea
\lefteqn{
\int_0^1 ds_1\ldots\int_0^1 ds_p\ \prod_{q=1}^\al
(s_1 s_2\ldots s_{q-1})} & & \nonumber \\
 & & \times\prod_{\al+1}^p
(1+s_{q-1}+s_{q-1}s_{q-2}+\cdots+s_{q-1}s_{q-2}\ldots s_2)\ .
\eea
Now for given $s_1,\ldots,s_\al$ we compute
\be
\int_0^1 ds_{\al+1}\ldots\int_0^1 ds_p
\prod_{\al+1}^p
(1+s_{q-1}+s_{q-1}s_{q-2}+\cdots+s_{q-1}s_{q-2}\ldots s_2)
\ee
by changing to the variables $y_{\al+1},\ldots,y_p$ defined by
\be
y_q\eqdef s_q
(1+s_{q-1}+s_{q-1}s_{q-2}+\cdots+s_{q-1}s_{q-2}\ldots s_2)
\ee
for $\al+1\le q\le p$. We then obtain, with $y_\al\eqdef
s_\al+s_\al s_{\al-1}+\cdots+ s_\al s_{\al-1}\ldots s_2\le\al-1$
\bea
\lefteqn{
\int_0^{1+y_\al}dy_{\al_+1}
\int_0^{1+y_{\al+1}}dy_{\al_+2}\ldots
\int_0^{1+y_{p-1}}dy_{p}} & & \nonumber \\
 & & \le\int_0^{1+y_\al}dy_{\al_+1}
\int_0^{1+y_{\al+1}}dy_{\al_+2}\ldots
\int_0^{1+y_{p-2}}e^{y_{p-1}}
dy_{p-1}
\label{input}\\
 &  & \le\int_0^{1+y_\al}dy_{\al_+1}
\int_0^{1+y_{\al+1}}dy_{\al_+2}\ldots
\int_0^{1+y_{p-3}}(e^{1+y_{p-2}}-1)
dy_{p-2} \\
 &  & \le e \int_0^{1+y_\al}dy_{\al_+1}
\int_0^{1+y_{\al+1}}dy_{\al_+2}\ldots
\int_0^{1+y_{p-3}}e^{y_{p-2}}
dy_{p-2} \label{output}
\eea
and, by repeating the argument leading from (\ref{input}) to
(\ref{output}), we get the inequality
\bea
\lefteqn{
\int_0^1 ds_{\al+1}\ldots\int_0^1 ds_p
\prod_{\al+1}^p
\lp\sum_{j=2}^{q}\prod_{j\le k \le q-1}s_k\rp} & & \nonumber \\
 & & \le e^{y_\al}\cdot e^{p-1-\al} \\
 &  & \le e^{\al-1}\cdot e^{p-1-\al}=e^{p-2}\ .
\eea
Therefore
\bea
\lefteqn{
\sum_{\si|J} \int_{1>h_1>\cdots>h_p>0} dh_1\ldots dh_p
\ \prod_{q=1}^p \frac{1}{h_{\si(q)}}} & & \nonumber \\
 &  &  \le e^{p-2}\times
\int_0^1 ds_1\ldots\int_0^1 ds_\al  \prod_{q=1}^\al
(s_1 s_2\ldots s_{q_1}) \\
&  & \le \frac{e^{p-2}}{\al!}
\eea
which proves the lemma.
\endproof

 Now the end of the proof of convergence is trivial:
\bea
\sum_\Gg \cB(\Gg) & \le & \sum_{p\ge 0}\sum_{J\subset\{1,\ldots,p\}}
\sum_{\si| J} K_8(n)K_9^p\la^p K_{10}^p n^{\#(J)} \nonumber \\
&  & \times  \int_{1>h_1>\cdots>h_p>0} dh_1\ldots dh_p
\prod_{q=1}^p \frac{1}{h_{\si(q)}} \\
& \le &  \sum_{p\ge 0}\sum_{0\le j\le p} \lp
\begin{array}{c} p \\ j
\end{array}
\rp
K_8(n)K_9^p\la^p K_{10}^p e^{p} \frac{n^j}{j!} \\
& \le &  K_8(n)  e^{n} \sum_{p\ge 0}
(2eK_9 K_{10}\la)^p<+\infty
\eea
for $\la$ small enough.
\endproof

\noindent{\bf Acknowledgments}
\medskip

We thank C. de Calan for his contribution to Lemma 9.

\end{document}